\documentclass[twocolumn,showpacs,amsmath,amssymb,prl]{revtex4-1}
\usepackage{graphicx}% Include figure files
\usepackage{dcolumn}% Align table columns on decimal point
\usepackage{bm}% bold math
\usepackage[abs]{overpic}
\usepackage{here}
\usepackage{color}
\def\up{\uparrow}
\def\down{\downarrow }
\def\Vec#1{\bm{#1}}

\begin{document}

%\preprint{}

\title{Inverse coherence effects in nuclear magnetic relaxation rates as a sign of topological superconductivity}

\author{Yuki Nagai}
\affiliation{CCSE, Japan  Atomic Energy Agency, 178-4-4, Wakashiba, Kashiwa, Chiba, 277-0871, Japan}

\author{Yukihiro Ota}
\affiliation{CCSE, Japan  Atomic Energy Agency, 178-4-4, Wakashiba, Kashiwa, Chiba, 277-0871, Japan}

\author{Masahiko Machida}
\affiliation{CCSE, Japan  Atomic Energy Agency, 178-4-4, Wakashiba, Kashiwa, Chiba, 277-0871, Japan}

\date{\today}% It is always \today, today,
             %  but any date may be explicitly specified
             
\begin{abstract}
We reveal that three-dimensional multi-orbital topological superconductivity can be identified by a bulk measurement, i.e., the temperature dependence of nuclear magnetic relaxation
 (NMR) rates.   
Below a critical temperature $T_{\rm c}$, 
the NMR rate in the topological state exhibits an anti-peak profile, which is opposite to the conventional $s$-wave state.
This inversion coherence effect  comes from a twist of order
 parameters with respect to orbital and spin degrees of freedom. 
Our self-consistent calculations in the model for
 Cu$_{x}$Bi$_{2}$Se$_{3}$ prove that the inverse coherence effect appears
 as a concave temperature dependence of the NMR rates. 
 We propose that a time-reversal-invariant orbital-singlet spin-triplet topological superconductivity 
% topologically non-trivial superconductivity 
is characterized by the 
 temperature dependence of the NMR rate. 
\end{abstract}

% PACS, the Physics and Astronomy
                             % Classification Scheme.
%\keywords{Suggested keywords}%Use showkeys class option if keyword
                              %display desired
\maketitle

Physics and topology have mutual connections in modern
physics~\cite{Nakahara,Fradkin}.  
Topological materials, such as topological insulators and
superconductors open an intriguing avenue of materials science and
quantum engineering~\cite{Hasan,Beenakker}. 
A reliable way of detecting these materials prompts the study
of topology in physics. 
Both surface and bulk probes are crucial for identifying topological materials. 
The bulk-boundary correspondence~\cite{Fradkin}
indicates that gapless surface states between different
topological states signal nontrivial topological order. 
Bulk quantities also contain a signature 
of their topological states. 
A definite example is the conductance in integer quantum Hall systems;
the quantized behaviors ruled by a topological invariant are
observed~\cite{QHE}. 
%Thus, 

Topological superconductors have attracted intense attention. 
A classification study with different kinds of spatial dimension
and symmetry~\cite{Schnyder:2008ez} indicates the presence of a
topological superconductivity in three-dimensional (3D) time-reversal-invariant
systems. 
Superconducting topological insulators, such as 
$\mbox{Cu}_{x}\mbox{Bi}_{2}\mbox{Se}_{3}$~\cite{Hor_L10,Wray_NP10,MKR_L11,Sasaki,Kirzhner}
and $\mbox{Sn}_{1-x}\mbox{In}_{x}\mbox{Te}$~\cite{SasakiSn}  
are the candidates of the bulk 3D topological
superconductors. 
Their topology is argued by $Z_{2}$ invariants~\cite{Fu2010,Sato:2010bi}. 
A mean-field model of
$\mbox{Cu}_{x}\mbox{Bi}_{2}\mbox{Se}_{3}$~\cite{Fu2010} leads to notable
predictions, such as vortex bound states~\cite{NagaiMajo,Nagai:2014cs} and
impurity effects~\cite{MiFu,Nagaiimp,Nagai:2015fv}. 
Among these studies, although the  gapless bound states on surfaces at low
temperatures are actively
studied~\cite{Sasaki,Kirzhner,Mizushima2014,NagaiQuasi,Hao;Lee:2015}, 
observing an intrinsic behavior via bulk measurements is rarely argued. 
In addition, the setups of surface measurements would be sensitively affected by the characteristics of the interfaces between materials and probes. 
Thus, a bulk measurement sensitive to topological characters is
important for supporting results in surface measurements and revealing
the properties of Cooper-pair condensations. 

A key quantity of connecting topological characters with bulk measurements is a correlation function. 
Current-current correlation functions, for example, lead to the quantized conductance in integer quantum Hall effects~\cite{QHE}. 
Spin-spin correlation functions are essential for superconductivity since these quantities well reflect the spin-state properties of Cooper pairs. 
Nuclear magnetic relaxation (NMR) rates ($T_{1}^{-1}$) are directly related to the spin-spin correlation functions. 
Therefore, it is interesting and important question to ask whether NMR rates contain any characteristic information on topological superconductivity. 
Specifically, the coherence effect is notable. 
The NMR rate in the presence of a spin-singlet $s$-wave superconducting state is enhanced just below a critical temperature $T_{\rm c}$, owing
to the coherence factor~\cite{Tinkham:2004}. 
This coherence peak (Hebel-Slichter peak) \cite{Hebel:1959eo,Masuda:1962bb} comes from the formation of superconducting gaps. 
The absence of the peak and a power-law behavior of $T_{1}^{-1}$ at low temperatures indicate the occurrence of unconventional states \cite{Bang:2009fl,Nagai:2008fr,Sigrist1991}. 

In this paper we claim that the bulk measurements of NMR rates detect 
a 3D odd-parity fully-gapped
topological superconducting state in time-reversal-invariant multi-orbital systems.
An {\it inverse} coherence effect just below $T_{\rm c}$ is the
signature of this odd-parity state; 
the coherence factor contributes to the NMR rates with an opposite sign
to that of the conventional $s$-wave states. 
Our self-consistent calculations in the model of 
$\mbox{Cu}_{x}\mbox{Bi}_{2}\mbox{Se}_{3}$~\cite{Fu2010} show that this
sign reversal leads to an anti-peak behavior of the NMR rates below
$T_{\rm c}$. 
Using  the Fermion anticommutation property, we show that the odd
parity allows a twist of a gap function with respect to
the internal degrees of freedom (e.g., orbital and spin) in systems,
causing the novel temperature dependence of NMR rates.
Hence, NMR rates are bulk quantities sensitive
to parity closely related to defining $Z_{2}$
invariants~\cite{Fu2010,Sato:2010bi} in a superconducting state.

We formulate the NMR rate of a multi-orbital superconductor, with a mean-field approach. 
The mean-field Bogoliubov-de Gennes (BdG) Hamiltonian is 
\(
{\cal H} 
= (1/2)
\sum_{\Vec{k}} 
\Vec{\psi}^{\dagger}_{\Vec{k}} \check{H}(\Vec{k}) \Vec{\psi}_{\Vec{k}}
\), 
with 
\(
\Vec{\psi}_{\Vec{k}}^{\dagger} 
= 
(\Vec{c}_{\Vec{k}}^{\dagger},\, \Vec{c}_{-\Vec{k}}^{\rm T})
\) 
and 
\(
\Vec{\psi}_{\Vec{k}}^{\rm T}
=
(\Vec{c}_{\Vec{k}}^{\rm T},\, \Vec{c}_{-\Vec{k}}^{\dagger} )
\). 
The $2n_{\rm o}$-component column (raw) vector 
\(
\Vec{c}_{\Vec{k}}
\) 
(\(
\Vec{c}_{\Vec{k}}^{\dagger}
\)) 
contains electron's annihilation (creation) operators, with the number
of orbitals $n_{\rm o}$. 
When $n_{\rm o}=2$, we have 
\(
\Vec{c}^{\rm T}
=
(c^{1}_{\uparrow},c^{1}_{\downarrow},
c^{2}_{\uparrow},c^{2}_{\downarrow})
\). 
The BdG matrix is 
\begin{align}
\check{H}(\Vec{k}) &= \left(\begin{array}{cc}
\hat{H}_{0}(\Vec{k}) & \hat{\Delta}(\Vec{k})\\
\hat{\Delta}^{\dagger}(\Vec{k}) & -[\hat{H}_{0}(-\Vec{k})]^{\ast}
\end{array}\right).
\end{align}
The normal-state Hamiltonian is $\hat{H}_{0}$. 
The pairing potential
(matrix) fulfills 
\(
 \hat{\Delta}^{\rm T}(- \Vec{k}) = - \hat{\Delta}(\Vec{k})
\), owing to the Fermion anticommutation property.
The check symbol ($\check{~}$) indicates a matrix in the Nambu space,
whereas the hat symbol ($\hat{~}$) does that in an orbital-spin space. 
The NMR rate~\cite{Hayashi:2006ky,Nagai:2008fr} is
\begin{align}
&\frac{1}{T_{1}(T) T} 
=  
\pi  \sum_{\alpha, \alpha^{\prime}} 
\int_{-\infty}^{\infty} d \omega \,
\left[
- \frac{d f(\omega)}{d \omega} 
\right]
\nonumber \\
& \times {\rm Re} \: 
\left\{ 
\rho_{\up \up}^{G \alpha \alpha^{\prime}}(\omega) 
\rho_{\down \down}^{G  \alpha^{\prime} \alpha}(\omega) 
-
\rho_{\up \down}^{F \alpha \alpha^{\prime}}(\omega) 
[\rho_{\down \up}^{F \alpha \alpha^{\prime}}(\omega)]^{\ast} 
\right\}. \label{eq:t1}
\end{align}
We use the unit system of $\hbar=k_{\rm B}=1$. 
The indices $\alpha$ and $\alpha^{\prime}$ represent orbital labels. 
The Fermi-Dirac distribution function is denoted by $f(\omega)$. 
The spectral functions $\hat{\rho}^{G}(\omega)$ and 
$\hat{\rho}^{F}(\omega)$ are the submatrices of 
\(
\check{\rho}^{G}(\omega) 
= 
(-1/2\pi i)
\sum_{\Vec{k}} 
\left[ 
\check{G}_{\Vec{k}}(i\omega_{n} \rightarrow \omega + i 0) 
- 
\check{G}_{\Vec{k}}(i\omega_{n} \rightarrow \omega - i 0)  
\right]
\). 
Temperature Green's function is 
\(
\check{G}_{\Vec{k}}(i\omega_{n})
=
[i\omega_{n} - \check{H}(\Vec{k})]^{-1}
\), with the fermionic Matsubara frequency $\omega_{n}=\pi T(2n+1)$ 
($n\in \mathbb{Z}$). 
The matrix form in the Nambu space is  
\begin{align}
 \check{G}_{\Vec{k}}(i\omega_{n}) 
&= 
\left(\begin{array}{cc}
 \hat{G}_{\Vec{k}}(i\omega_{n}) & \hat{F}_{\Vec{k}}(i\omega_{n}) \\
 \hat{\bar{F}}_{\Vec{k}}(i\omega_{n}) &  \hat{\bar{G}}_{\Vec{k}}(i\omega_{n})
\end{array}\right).
\end{align}
The diagonal block, $\hat{G}_{\Vec{k}}$ leads to $\hat{\rho}^{G}$, relevant to electron's density of states.
The off-diagonal block, $\hat{F}_{\Vec{k}}$ contributes to the anomalous spectral function $\hat{\rho}^{F}$.

The presence of a coherence peak just below $T_{\rm c}$ is
determined by the second term in Eq.~(\ref{eq:t1})~\cite{Hayashi:2006ky,Sigrist1991}. 
The Hebel-Slichter peak appears when this term (including the
minus sign in front of the spectral functions) has a positive
contribution to $T_{1}^{-1}$. 
To intuitively understand the behaviors of the second term, we
evaluate anomalous Green's functions near $T_{\rm c}$. 
Linearizing $\check{G}$ with respect to
$\hat{\Delta}$, we obtain 
\(
\sum_{\Vec{k}} \hat{F}_{\Vec{k}}(i \omega_{n}) 
\approx 
\sum_{\Vec{k}} 
\hat{G}^{\rm N}_{\Vec{k}}(i \omega_{n}) 
\hat{\Delta}(\Vec{k}) 
\hat{\bar{G}}^{\rm N}_{\Vec{k}}(i \omega_{n})
\), with normal-state Green's functions, $\hat{G}_{\Vec{k}}^{\rm N}$
and $\hat{\bar{G}}_{\Vec{k}}^{\rm N}$. 
The spin-singlet property of $s$-wave states indicates 
\(
\Delta_{\uparrow \downarrow} =
-\Delta_{\downarrow \uparrow}
\); 
we obtain 
\(
-\rho^{F}_{\uparrow \downarrow} (\rho^{F}_{\downarrow \uparrow})^{\ast}
=
+|\rho^{F}_{\uparrow \downarrow}|^{2}
\), with $\hat{\rho}^{F} \neq 0$. 
In contrast, the $\Vec{k}$-integral with a spin-triplet $p$-wave gap
leads to $\hat{\rho}^{F}=0$ \cite{Hayashi:2006ky}. 

Now, we apply the above arguments to topological superconducting insulators. 
We focus on the model of \(\mbox{Cu}_{x}\mbox{Bi}_{2}\mbox{Se}_{3}\).  
The normal electrons are
effectively described by a two-orbital model ($n_{\rm o} = 2$),
leading to a massive Dirac Hamiltonian, 
\begin{align}
\hat{H}_{0}(\Vec{k}) 
&= 
\gamma^{0}\bigg[
-\mu \Gamma^{0}
+
\sum_{i=1}^{3} v_{i} k_{i} \Gamma^{i}
+
m \Gamma^{4}
+
h_{5}(\Vec{k})\Gamma^{5}
\bigg], 
\label{eq:normalH}
\end{align}
with chemical potential $\mu$, spin-orbit coupling constants $v_{i}$,
and mass $m$.  
This Hamiltonian is the same as that in Ref.~\cite{Fu2010}, except
$h_{5}$. 
The last term corresponds to the effects of hexagonal warping in the
Fermi surface of
$\mbox{Cu}_{x}\mbox{Bi}_{2}\mbox{Se}_{3}$~\cite{Fu:2014dc}. 
We drop this term in most of our calculations, but argue
the effects on the NMR rate at the end of this paper. 
Six kinds of $4\times 4$ matrices $\Gamma^{A}$
($A=0\,,1\,\ldots,5$) are composed of the gamma matrices  
$\gamma^{\mu}$ ($\mu=0,\,1,\,2,\,3$)\cite{Peskin} and the identity: 
\(
\Gamma^{A} = \gamma^{A}
\) 
($A \neq 4$)
and 
\(
\Gamma^{4} = \openone_{4}
\), 
with $\gamma^{5} = i\gamma^{0}\gamma^{1}\gamma^{2}\gamma^{3}$. 
Our choice~\cite{note} is that 
\(
\gamma^{0} = \sigma_{x} \otimes \openone_{2}
\), 
\(
\gamma^{1} = -i \sigma_{y} \otimes s_{y}
\),
\(
\gamma^{2} = i \sigma_{y} \otimes s_{x}
\),
and
\(
\gamma^{3} = i \sigma_{z} \otimes \openone_{2}
\), 
where 
$\sigma_{x,\,y,\,z}$ ($s_{x,\,y,\,z}$) are the $2 \times 2$ Pauli
matrices in the orbital (spin) space. 
In this paper, we study the momentum-independent pair potential $\hat{\Delta}$, owing to 
the onsite interaction~\cite{Fu2010,Sato:2010bi}.
The relation $\Delta^{\rm T}=-\Delta$ leads to 
\begin{align}
\hat{\Delta} 
&= \sum_{A=0}^{5} \Delta^{A} \Gamma^{A} \gamma^{2} \gamma^{5}. 
\end{align}
Since $\gamma^{2}\gamma^{5} = \openone_{2}\otimes s_{y}$, the case of
$\Gamma^{A}$ to be the identity (i.e., $A=4$) describes a spin-singlet
$s$-wave state: 
\(
\Delta^{4}
\propto
\sum
\langle 
c_{-\Vec{k} \downarrow}^{1}c_{\Vec{k} \uparrow}^{1}
+
c_{-\Vec{k} \downarrow}^{2}c_{\Vec{k} \uparrow}^{2}
\rangle
\). 
The additional $\Gamma^{A}$ ($A \neq 4$) characterizes 
a \textit{twist} of each order-parameter component in the orbital-spin
space, compared to the conventional $s$-wave state. 
According to Ref.~\cite{Fu2010}, even-parity order parameters ($A_{1g}$
states) are given by $\Delta^{4}$ and $\Delta^{0}$. 
Odd-parity states correspond to $\Delta^{1,2,3,5}$. 
The component $\Delta^{5}$ corresponds to an odd-parity
fully-gapped ($A_{1u}$) state~\cite{Fu2010,Sato:2010bi}: 
\(
\Delta^{5}
\propto
\sum 
\langle
c_{-\Vec{k} \downarrow}^{2}c_{\Vec{k} \uparrow}^{1}
+
c_{-\Vec{k} \uparrow}^{2}c_{\Vec{k} \downarrow}^{1}
\rangle
\). 
The odd-parity state on Dirac-type normal electrons given by Eq.~(\ref{eq:normalH}) \cite{Newnote} is 
nontrivial, in terms of not only a $Z_{2}$ invariant\cite{Fu2010,Sato:2010bi}, but also a 1D $Z_{2}$ invariant \cite{Sato:2010bi,note1d}. 

Let us study the anomalous spectral function $\hat{\rho}^{F}$
of the odd-parity state near $T_{\rm c}$.
Normal-state Green's functions are evaluated by an algebraic relation of
$\hat{H}_{0}$; we find that 
\(
[\hat{H}_{0}^{\prime}(\Vec{k})]^{2} 
= [E(\Vec{k})]^{2} 
\), 
with 
\(
\hat{H}_{0}^{\prime} 
= 
\hat{H}_{0} + \mu
\). 
This property corresponds to the fact that the Dirac equation is the
square root of the Klein-Gordon equation\cite{Peskin}.  
Hence, we
obtain 
\(
\hat{G}^{\rm N}_{\Vec{k}}( i\omega_{n})
=
\sum_{\ell=\pm}
\hat{P}_{\ell}(\Vec{k})/[i \omega_{n} - \ell E(\Vec{k}) + \mu]
\), with 
\(
\hat{P}_{\pm}
= 
(1/2)[1 \pm (\hat{H}_{0}^{\prime}/E)]
\). 
The projectors $\hat{P}_{\ell}$ are rewritten by 
\(
\hat{P}_{\ell}(\Vec{k})
= 
\gamma^{0} 
\sum_{A=0}^{5} w_{\ell}^{A}(\Vec{k}) \Gamma^{A}
\). 
Then, when 
\(
\hat{\Delta}(\Vec{k}) 
= \hat{\Delta}^{5} 
\equiv  \Delta^{5} \Gamma^{5}\gamma^{2}\gamma^{5}
\) and $h_{5}=0$, we find that 
\begin{align}
 \hat{\rho}^{F}(\omega; \Delta^{5}) 
&=  \chi^{5}(\omega) \hat{\Delta}^{5}
+ \chi^{40}(\omega) \gamma^{0} \hat{\Delta}^{5},\label{eq:rhof}
\end{align}
near $T_{\rm c}$. 
The coefficients are 
\(
\chi^{5,\,40}
= 
\sum_{\Vec{k}} \sum_{\ell,\ell^{\prime}} 
N_{\ell,\ell^{\prime}}(\Vec{k},\omega)
{\cal W}_{\ell,\ell^{\prime}}^{5,\,40}(\Vec{k})
\), 
where
\(
N_{\ell,\ell^{\prime}}
= 
\{ 1/[(\ell + \ell^{\prime}) E - 2 \mu \}
[
\delta(\omega - \ell E + \mu )
-
\delta(\omega + \ell^{\prime} E - \mu )
]
\),
\(
{\cal W}_{\ell,\ell^{\prime}}^{5}
=
\sum_{A=0}^{3} 
w_{\ell}^{A}  w_{\ell^{\prime}}^{A}
- 
w_{\ell}^{4}  w_{\ell^{\prime}}^{4}
\)
and 
\(
\mathcal{W}^{40}_{\ell,\ell^{\prime}}
=
w_{\ell}^{4} w^{0}_{\ell^{\prime}}
-
w_{\ell}^{0} w^{4}_{\ell^{\prime}}
\). 
Since $\Gamma^{5}\gamma^{2}\gamma^{5} = -i \sigma_{y} \otimes s_{x}$, 
$\hat{\Delta}^{5}$ is \textit{odd} under an orbital-index exchange, while
is {\it even}  under a spin-index exchange. 
The multiplication of $\gamma^{0}$ with $\hat{\Delta}^{5}$ does not
change the property of a spin-index exchange.  
Thus, we show that 
\(
 \rho_{\up \down}^{F \alpha \alpha^{\prime}}(\omega) 
=  \rho_{\down \up}^{F \alpha \alpha^{\prime}}(\omega)
\). 
This property indicates that the second term in
Eq.~(\ref{eq:t1}) \textit{negatively} contributes to $T_{1}^{-1}$: 
\(
-
\rho^{F \alpha \alpha^{\prime}}_{\uparrow \downarrow}
(\rho^{F \alpha \alpha^{\prime}}_{\downarrow \uparrow})^{\ast}
=
-
|\rho^{F \alpha \alpha^{\prime}}_{\uparrow \downarrow} |^{2}
\). 
An inverse coherence effect can occur
 in the
presence of the odd-parity state. 

\begin{figure}[t]
%%%%--- I comment out figure regions
%\vspace{50mm}
\begin{center}
     \begin{tabular}{p{ 0.8 \columnwidth}} %p{0.5 \columnwidth}}%  p{28mm}}
      \resizebox{0.8 \columnwidth}{!}{\includegraphics{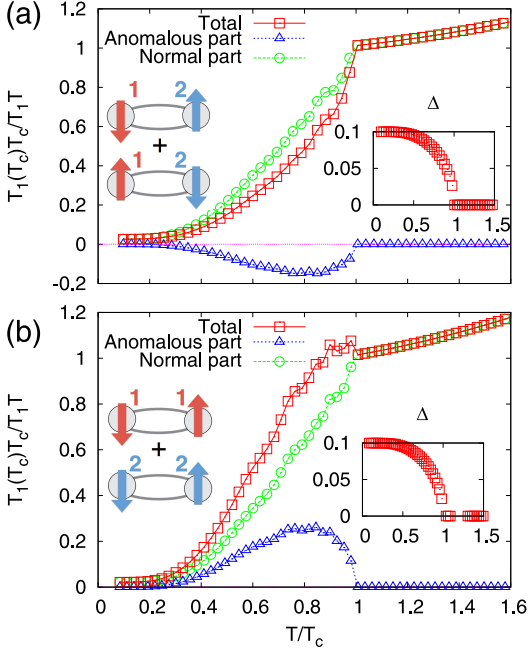}} 
      %\\ %&
     % (b)\resizebox{0.8 \columnwidth}{!}{\includegraphics{Fig1b.eps}} 
    \end{tabular}
\end{center}
\caption{
\label{fig:fig1}(Color online)
Temperature dependence of nuclear magnetic relaxation rates in (a) an odd-parity gap 
\(
\Delta^{5}
\propto
\sum 
\langle
c_{-\Vec{k} \downarrow}^{2}c_{\Vec{k} \uparrow}^{1}
+
c_{-\Vec{k} \uparrow}^{2}c_{\Vec{k} \downarrow}^{1}
\rangle
\)
(
$A_{1u}$ \cite{Fu2010} or $\Delta_{2}$ \cite{Sasaki}
)
and (b) 
an even-parity gap 
\(
\Delta^{4}
\propto
\sum
\langle 
c_{-\Vec{k} \downarrow}^{1}c_{\Vec{k} \uparrow}^{1}
+
c_{-\Vec{k} \downarrow}^{2}c_{\Vec{k} \uparrow}^{2}
\rangle
\) 
($A_{1g}$ \cite{Fu2010} or $\Delta_{1a}$ \cite{Sasaki}
). 
We set the chemical potential $\mu = 0.8$ and the Dirac mass $m=0.4$ [See, Eq.~(\ref{eq:normalH})]. 
We tune either $U$ or $V$ so that the zero-temperature gap amplitude takes $0.01$.  
Inset: temperature dependence of the superconducting gaps. }
\end{figure}

Now, we show the inverse coherence effect of $T_{1}^{-1}$, solving the BdG equations self-consistently. 
We focus on Eq.~(\ref{eq:normalH}) with $h_{5}=0$. 
We set $v_{i} = v$, for simplicity. 
We take a unit system with $v = 1$. 
We vary $m$, with fixed chemical potential $\mu = 0.8$. 
The superconducting pairing model is two-orbital on-site
density-density interaction~\cite{Fu2010,Mizushima2014}, 
$H_{\rm int}(\Vec{x}) 
= 
U \left[ n_{1}(\Vec{x}) ^{2} + n_{2}(\Vec{x}) ^{2} \right] 
+2 V n_{1}(\Vec{x})  n_{2}(\Vec{x})$ 
with 
\(
n_{\alpha}
= 
\sum_{s=\up,\down}  c_{s}^{\alpha\, \dagger} c_{s}^{\alpha}
\). 
The gap equation is 
\begin{align}
\Delta_{a a^{\prime}} 
&= 
- {\cal V}_{a a^{\prime}} \, 
n^{F}_{a a^{\prime}}, 
\label{eq:gap}
\end{align}
with 
\(
\hat{n}^{F}
= -  
\sum_{\Vec{k}}  
\lim_{\tau \rightarrow 0+} 
\exp(- i \omega_{n} \tau ) \hat{F}_{\Vec{k}}(i \omega_{n})
\)
and  
\(
a = (\alpha,s)
\). 
The intra-orbital coupling constant $U$ and the inter-orbital coupling
constant $V$ are bundled up by $\mathcal{V}_{a a^{\prime}}$, i.e., 
\(
{\cal V}_{a a^{\prime}} = U
\)
with $\alpha = \alpha^{\prime}$
and
\(
{\cal V}_{a a^{\prime}} = V
\) 
with $\alpha \neq \alpha^{\prime}$. 
Self-consistently solving Eq.~(\ref{eq:gap}), we evaluate the
temperature dependence of Eq.~(\ref{eq:t1}). 
The $\Vec{k}$ integrals are performed by the trapezoidal rule in the
spherical coordinate system, with cutoff momentum $k_{\rm max}=9$ and
mesh $(N_{k},N_{\theta},N_{\phi}) = (128,128,128)$. 
The smearing factor of the delta function is set by $0.01$. 
We tune either $U$ or $V$ so that the zero-temperature gap amplitude takes $0.01$. 
In this paper, we focus on the odd-parity gap function 
\(
\Delta^{5}
\propto
\sum 
\langle
c_{-\Vec{k} \downarrow}^{2}c_{\Vec{k} \uparrow}^{1}
+
c_{-\Vec{k} \uparrow}^{2}c_{\Vec{k} \downarrow}^{1}
\rangle
\) 
and the even-parity gap function 
\(
\Delta^{4}
\propto
\sum
\langle 
c_{-\Vec{k} \downarrow}^{1}c_{\Vec{k} \uparrow}^{1}
+
c_{-\Vec{k} \downarrow}^{2}c_{\Vec{k} \uparrow}^{2}
\rangle
\). 
Assessing the linearized gap equation~\cite{Fu2010}, we find that the
former appears as the most stable phase when $U=0$ and
$V<0$ (attractive inter-orbital interaction). 
The latter occurs when $U<0$ and $V=0$ (attractive
intra-orbital interaction). 

\begin{figure}[th]
\begin{center}
     \begin{tabular}{p{ 0.8 \columnwidth}} %p{0.5 \columnwidth}}%  p{28mm}}
      \resizebox{0.8 \columnwidth}{!}{\includegraphics{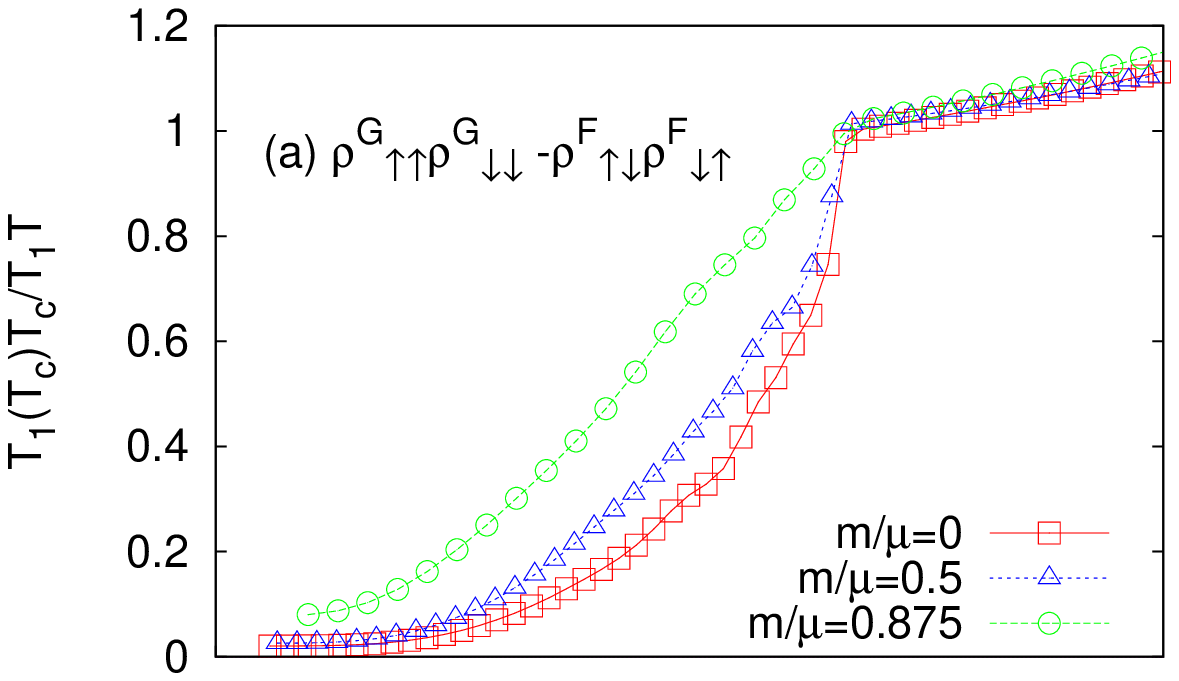}} \\ %&
     \resizebox{0.8 \columnwidth}{!}{\includegraphics{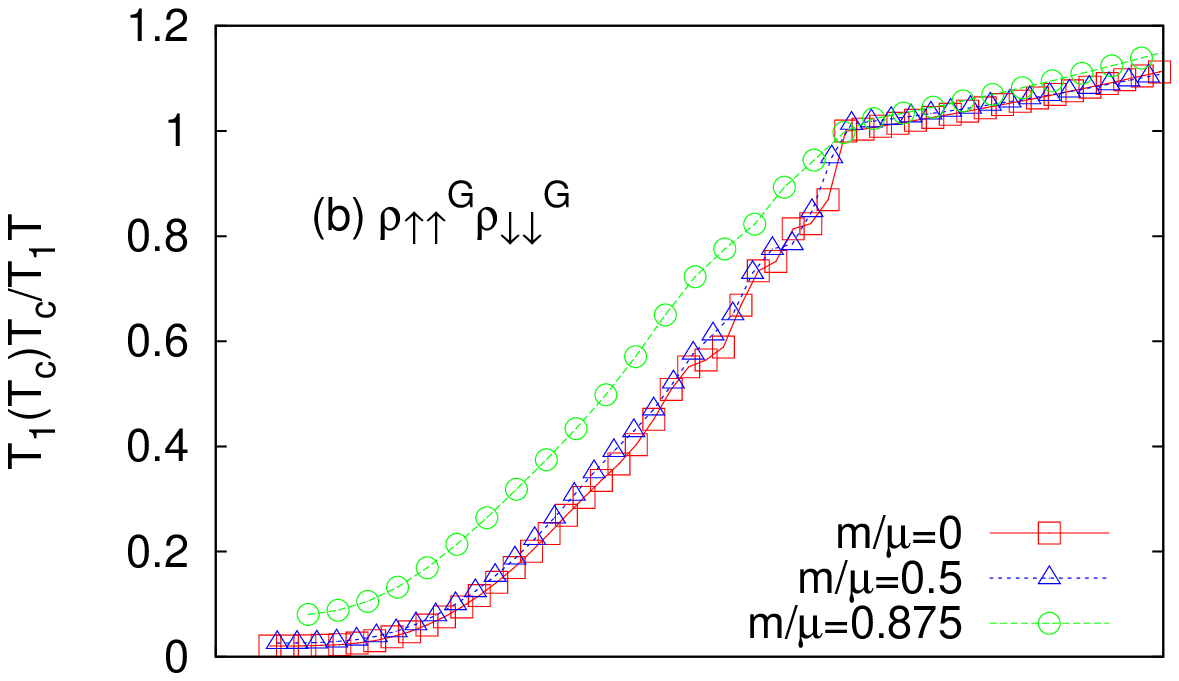}} \\
     \resizebox{0.8 \columnwidth}{!}{\includegraphics{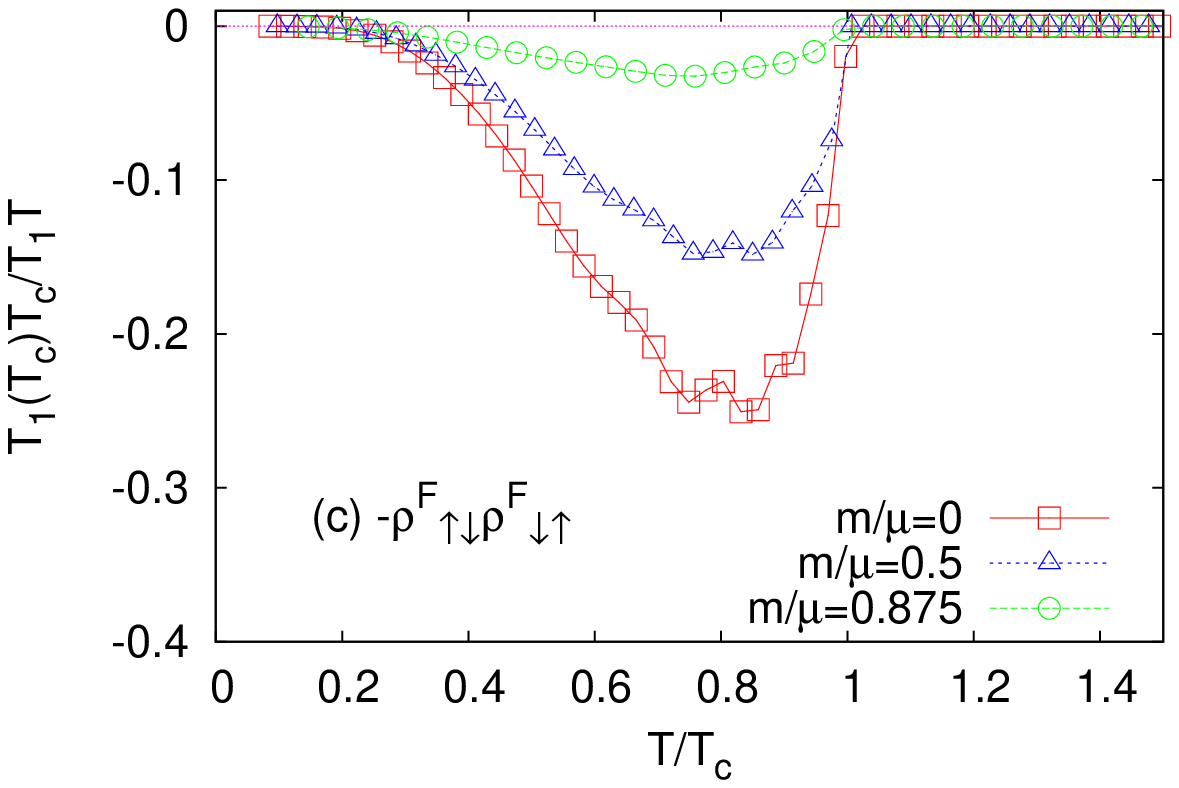}} 
    \end{tabular}
\end{center}
\caption{\label{fig:fig2}(Color online)
Nuclear magnetic relaxation rates of an odd-parity gap $\Delta^{5}$ (
$A_{1u}$ \cite{Fu2010} or $\Delta_{2}$ \cite{Sasaki}
), changing the Dirac mass $m$ with fixed chemical potential $\mu = 0.8$.
The total rate (a), the normal part (b), and the anomalous part (c) are shown. 
}
\end{figure}

Figure \ref{fig:fig1} shows the temperature dependence of 
$(T_{1}T)^{-1}$, with $m=0.4$. 
When the gap function is odd parity [Fig.~\ref{fig:fig1}(a)], an anti-peak or concave behavior occurs below $T_{\rm c}$, coming from the anomalous
part of $(T_{1}T)^{-1}$ (blue triangles). 
The temperature dependence is contrast to that of
Fig.~\ref{fig:fig1}(b), in which the gap function is even parity 
and the coherence peak appears just below $T_{\rm c}$. 
Thus, we obtain the inverse coherence effect in
$\hat{\Delta}^{5}$. 
It is worth noting that the temperature dependence of the gap function
(insets in Fig.~\ref{fig:fig1}) has no significant difference between
$\hat{\Delta}^{5}$ and $\hat{\Delta}^{4}$. 
Moreover, we can find that the exponential temperature dependence of
$(T_{1}T)^{-1}$ occurs at lower temperatures; this is consistent with the
fully-gaped feature of $\hat{\Delta}^{5}$. 
Thus, the temperature dependence of the NMR rate definitely signals the
emergence of $\hat{\Delta}^{5}$. 

\begin{figure}[bh]
\begin{center}
     \begin{tabular}{p{ 0.7 \columnwidth}} %p{0.5 \columnwidth}}%  p{28mm}}
      \resizebox{0.7 \columnwidth}{!}{\includegraphics{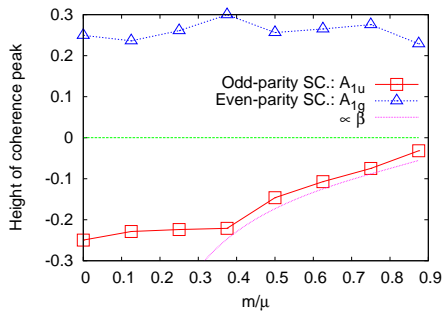}} 
    \end{tabular}
\end{center}
\caption{\label{fig:fig3}(Color online) 
Height of coherence peaks [normalized by the value at $T_{\rm c}$, $T_{1}(T_{\rm c})T_{\rm c}$], changing $m/\mu$ (Dirac mass/chemical potential). 
The negative height means the inverse coherence effects. 
}
\end{figure}

Let us examine how the inverse coherence effect depends on material
variables, focusing on mass $m$.  
Figure \ref{fig:fig2}(a) shows the $m$-dependence of $(T_{1}T)^{-1}$ when the
gap function is $\hat{\Delta}^{5}$. 
The concave behavior is much pronounced when $m$ is
small. 
The normal parts [Fig.~\ref{fig:fig2}(b)]  of the NMR rate do not considerably depend on $m$, while the anomalous one [Fig.~\ref{fig:fig2}(c)] goes to
zero when $m \to \mu$.
Let us evaluate the extreme value of the anomalous part below
$T_{\rm c}$ normalized by $T_{1}(T_{\rm c})T_{\rm c}$. 
This quantity characterizes the amount of the coherence
effect. 
Figure \ref{fig:fig3} shows the extrema (i.e., height), changing $m/\mu$. 
We find that when we take $\hat{\Delta}^{4}$, the extrema is
positive and almost independent of $m/\mu$. 
In contrast, in the case of $\hat{\Delta}^{5}$ the extrema is
negative and drastically depends on $m/\mu$. 
The value approaches zero when $m /\mu \to 1$, while it takes a
negative constant when $m /\mu \to 0$. 
When $m \sim \mu$, the odd-parity fully-gaped gap function is mapped to the spin-triplet $p$-wave gap function with
$d$-vector $\Vec{d} \propto \Vec{k}$\cite{Nagaiimp}. 
Thus, the inverse coherence effects are suppressed when
$m$ becomes large. 
We also find that the negative extrema as  $m/\mu \to 1$ is
proportional to an indicator of relativistic effects~\cite{Nagaiimp},  
\(
\beta = |v k_{\rm F}|/|m| = \sqrt{(\mu/m)^{2}-1}
\), where $k_{\rm F}$ is the Fermi momentum. 
It indicates that a relativistic effect, which is intrinsic in the
Dirac Hamiltonian, characterizes the inverse coherence effect of the NMR
rate. 
We stress that the anti-peak behaviors are predominate when $m/\mu \to 0$. According to Refs.~\cite{Nagaiimp,Nagai:2015fv,MiFu}, the 
odd-parity state with small Dirac mass is robust against non-magnetic impurities. 
Thus, one can detect the inverse coherence effect even in dirty samples. 

\begin{table*}[t]
\caption{
Parity table of different gap functions.
The coherence effect  is characterized by 
spin parity $p_{\rm s}$ [$\Delta_{\up \down}^{\alpha \alpha^{\prime}}(\Vec{k}) = p_{\rm s} \Delta_{\down \up}^{\alpha \alpha^{\prime}}(\Vec{k})$], 
momentum parity $p_{\rm m}$ [$\Delta_{\up \down}^{\alpha \alpha^{\prime}}(\Vec{k}) = p_{\rm m} \Delta_{\up \down}^{\alpha \alpha^{\prime}}(-\Vec{k})$], and 
orbital parity $p_{\rm o}$ [$\Delta_{\up \down}^{\alpha \alpha^{\prime}}(\Vec{k}) = p_{\rm o} \Delta_{\up \down}^{\alpha^{\prime} \alpha}(\Vec{k})$], with $p_{\rm s} p_{\rm m} p_{\rm o} = -1$.  
}
\label{table:1}
\begin{ruledtabular}
\begin{tabular}{ccccccc}
Gap type &
Spin parity $p_{\rm s}$&
Momentum parity $p_{\rm m}$& 
Orbital parity $p_{\rm o}$& 
Coherence effect \\
2-orbital $s$-wave SC. ($A_{1g}$)& -1 & 
+1 &
+1 &
positive   \\
2-orbital topo. SC. ($A_{1u}$)& +1 & 
+1 &
-1 &
negative \\
1-orbital $s$-wave SC.& -1 &
+1  &
+1 &
positive   \\
1-orbital $p_{x}+i p_{y}$-wave SC.& +1 & 
-1 &
+1 &
0  
\end{tabular}
\end{ruledtabular} 
\end{table*}

Now, we propose an unified point of view on NMR
rates near $T_{\rm c}$, depending on different gap
functions. 
We focus on three kinds of parity $p_{\rm s}$, $p_{\rm m}$ and 
$p_{\rm o}$ on $\hat{\Delta}$.  
Spin parity $p_{\rm s}$ is related to the exchange of spin: 
\(
\Delta_{s^{\prime} s}^{\alpha\alpha^{\prime}}(\Vec{k})
=
p_{\rm s}\Delta_{s s^{\prime}}^{\alpha\alpha^{\prime}}(\Vec{k})
\) with $p_{\rm s}= \pm 1$. 
Similarly, momentum (orbital) parity $p_{\rm m}(p_{\rm o})$ is determined by exchange of momentum (orbital): 
\(
\Delta_{s s^{\prime}}^{\alpha\alpha^{\prime}}(-\Vec{k})
=
p_{\rm m}\Delta_{s s^{\prime}}^{\alpha\alpha^{\prime}}(\Vec{k})
\)
[
\(
\Delta_{s s^{\prime}}^{\alpha^{\prime}\alpha}(\Vec{k})
=
p_{\rm o}\Delta_{s s^{\prime}}^{\alpha\alpha^{\prime}}(\Vec{k})
\)
] with $p_{{\rm m} (\rm {o})} = \pm 1$. 
The fundamental relation of $\hat{\Delta}$ is 
\(
\hat{\Delta}^{\rm T}(-\Vec{k}) = -\hat{\Delta}(\Vec{k})
\), leading to 
\(
 p_{\rm s} p_{\rm m} p_{\rm o} = -1 
\). 
They are not independent of parity; 
in the present model the odd orbital parity is forbidden in the even-parity
fully-gapped states~\cite{suppl}. 
Table I summarizes $(p_{\rm s}, \,p_{\rm m},\, p_{\rm o})$ of different
gaps. 
The non-zero contributions of $\hat{\rho}^{F}$ to $T_{1}^{-1}$ is
allowed when $p_{\rm m}$=1. 
In fast, a single-orbital spin-triplet $p$-wave
state ($p_{\rm m} = -1$) has no peak~\cite{Sigrist1991,Hayashi:2006ky}.  
The sign of the coherence effect is determined by the spin parity
$p_{\rm s}$.  
A multi-orbital odd-parity state allows both $p_{\rm m} = 1$ and 
$p_{\rm s} = 1$~\cite{suppl}. 

We argue the application of NMR rates to finding 3D topological
superconductors. 
The knowledge of normal-state Fermi surfaces is needed \cite{Sato:2010bi,suppl}, to pick up plausible samples in terms of $Z_{2}$ invariants. 
After such a screening test, NMR measurements would lead to one of the evidence for the odd-parity fully-gapped state. 
The details are summarized in Ref.~\cite{suppl}.

Finally, we discuss the NMR rate with anisotropic odd-parity states labeled by $E_{u}$: 
\(
\Delta^{1}
\propto
\sum
\langle 
c_{-\Vec{k} \uparrow}^{1}c_{\Vec{k} \uparrow}^{1}
+
c_{-\Vec{k} \downarrow}^{2}c_{\Vec{k} \downarrow}^{2}
\rangle
\) \cite{Fu:2014dc,Nagai:2015fv,Nagai:2014cs}. 
According to Ref.~\cite{Fu:2014dc}, this pairing can be induced in the presence of a hexagonal warping term \(
h_{5} = i\lambda (k_{+}^{3} + k_{-}^{3}) 
\) 
with $k_{\pm} \equiv k_{x} \pm i k_{y}$. 
Near 
$T_{\rm c}$, we can find that
\(
\rho^{F\, \alpha \alpha^{\prime}}_{\uparrow \downarrow}
(\omega; \Delta^{1})
 \propto \lambda
\).
Since the orbital parity is odd but the spin parity is even, the
inverse coherence effect can occur within $|\lambda| \neq 0$. 
We will conclusively discuss the coherence effect of the NMR rate in
topological superconductors elsewhere, including anisotropy and momentum
dependence in the gap functions. 

In summary, 
we showed a bulk measurement, the NMR rate near $T_{\rm c}$ as a sign of topological superconductivity. 
The inverse coherence effect was predicted in the NMR rate below $T_{\rm c}$. 
Our self-consistent calculations in the model of Cu$_{x}$Bi$_{2}$Se$_{3}$ lead to the temperature dependence of $(T_{1}T)^{-1}$, 
with a concave behavior below $T_{\rm c}$. 
This notable effect originates from the anomalous part of the NMR rate in the odd-parity gap 
\(
\Delta^{5}
\propto
\sum 
\langle
c_{-\Vec{k} \downarrow}^{2}c_{\Vec{k} \uparrow}^{1}
+
c_{-\Vec{k} \uparrow}^{2}c_{\Vec{k} \downarrow}^{1}
\rangle
\).
Moreover, the negative height when $m/\mu \rightarrow 1$ is proportional to an indicator of normal-electron relativistic effects. 
Therefore, we claim that the detection of the inverse coherence effect definitely signals the emergence of bulk topological superconductors. 
Our approach of focusing on correlation functions unveiled a link of bulk measurements with a topological superconducting state. 
Accordingly, the present study can be a clue of building an explicit formula of topological invariants, with aid of bulk-measurement quantities in superconductors.

Y.N. thanks L. Fu for helpful comments on the hexagonal warping term. 
The calculations were performed by the supercomputing 
system PRIMERGY BX900 at the Japan Atomic Energy Agency. 
This study was partially supported by JSPS KAKENHI Grant Number 26800197 and Grant-in-Aid for Scientific Research (S) Grant Number 23226019.

%\section*{references}

\clearpage
%\newpage
%\widetext
\onecolumngrid
\setcounter{equation}{0}
\renewcommand{\thefigure}{S\arabic{figure}} 

\setcounter{figure}{0}

\renewcommand{\thesection}{S\arabic{section}.} 
%\section{Supplemental Material}
\begin{flushleft} 
{\Large {\bf Supplemental material}}
\end{flushleft} 
%\vspace{2mm}
%\begin{flushleft} 
%{\bf S1. Topological invariant and its detection}
%\end{flushleft} 

We show a way of classifying time-reversal-invariant (TRI)
3D topological superconductors, in terms of three kinds of parity. 
The setting of physical model here is more general than that in the main
text.  
We impose that normal electrons are subjected to a two-orbital
Hamiltonian with time-reversal and spatial-inversion symmetries, 
and the inversion transformation, $\hat{P}$ is expressed by one of the
orbital Puali matrices, say, $\sigma_{x}$. 
Three kinds of parity are related to, respectively, orbital degrees,
spatial inversion, and Fermi-surface topology~\cite{Sato:2010bi}. 
The orbital parity of pairing potential $\hat{\Delta}(\Vec{k})$ is
defined by partial transpose with respect to orbital
indices, given by 
\(
\mathcal{T}_{\rm o}[\hat{\Delta}(\Vec{k}) ]
=
(1/2)
[
\hat{\Delta}(\Vec{k})
+
\sum_{i=x,y,z}
(\sigma_{i}\otimes \openone )
\hat{\Delta}(\Vec{k}) 
(\sigma_{i} \otimes \openone)^{\ast}
]
\). 
Thus, we have
\(
\mathcal{T}_{\rm o}[\hat{\Delta}(\Vec{k}) ]
=
p_{\rm o}
\hat{\Delta}(\Vec{k})
\), 
with $p_{\rm o} = \pm 1$. 
A straightforward calculation shows that this definition is equal to
that in the main text. 
The spatial-inversion parity of $\hat{\Delta}(\Vec{k})$ is defined by 
\(
\hat{P}^{\dagger} \hat{\Delta}(\Vec{k}) \hat{P}^{\ast}
=
p \hat{\Delta}(-\Vec{k})
\), with $p=\pm 1$~\cite{Sato:2010bi}. 
The parity of Fermi-surface topology is given by~\cite{Sato:2010bi}
\begin{align}
p_{\rm F}^{ij} 
= 
\prod_{b} 
{\rm sgn} [ \epsilon_{2 b}(\Gamma_{i})]\, 
{\rm sgn} [ \epsilon_{2 b} (\Gamma_{j})], 
\end{align}
where $\epsilon_{2b}(\Gamma_{i})$ is the eigenvalue of
normal-electron Hamiltonian $\hat{H}_{0}(\Gamma_{i})$ at TRI momentum
point $\Gamma_{i}$ in the first Brillouin zone. 
Since the system has time-reversal symmetry, the Kramers degeneracy at
$\Gamma_{i}$ indicates that 
$\epsilon_{2 b}(\Gamma_{i}) = \epsilon_{2 b+1}(\Gamma_{i})$. 
Thus, the subscript $b$ runs over half of labels in the eigenvalues
of $\hat{H}_{0}(\Vec{k})$. 
According to Ref.~\cite{Sato:2010bi}, within weak-coupling
superconductivity, a 1D $Z_{2}$ invariant of a TRI odd-parity state
is defined by 
\(
(-1)^{\tilde{\nu}[C_{ij}]} = p_{\rm F}^{ij}
\), with a closed path $C_{ij}$ connecting between $\Gamma_{i}$ and
$\Gamma_{j}$. 
If 
$(-1)^{\tilde{\nu}[C_{ij}]} = -1$, the system is topologically
nontrivial~\cite{Sato:2010bi}. 
Moreover, odd numbers of zero-energy states are predicted on the
boundaries made of opening up a closed path $C_{ij}$. 
Table \ref{tab:classification} shows all the eight patterns  
of $(p_{\rm o},\, p,\, p_{\rm F}^{ij})$ and the corresponding 1D $Z_{2}$
when $p=-1$. 
We show that, in a two-orbital system, cases V and V\!I are forbidden since $p_{\rm o}$ and $p$ are not
independent of each other~\cite{note:orbital_parity}. 

Now, we summarize a method of detecting TRI 3D odd-parity topological
superconductors, using the link of the classification table with the
inverse coherence effects in nuclear magnetic relaxation (NMR) rates. 
If the gap function is an odd function of momentum 
(e.g., $\propto k_{x}+ik_{y}$), no coherence peak is expectable. 
Therefore, we focus on the case of even momentum parity in Table I of
the main text; the inverse coherence peaks can appear when  
$p_{\rm o}=-1$.  
Cases  V\!I\!I and V\!I\!I\!I in 
table \ref{tab:classification} allow an anti-peak behavior in NMR
rates.  
We stress that, without empolying any NMR experiments in a
superconducting state, case V\!I\!I is distinguished from case
V\!I\!I\!I by the Fermi-surface topology attainable from
normal-state probes, such as the first-principles calculations
of energy band and angular-resolved photo-emission spectroscopy of Fermi
surfaces. 
After a screening test of Fermi surface topology, NMR rate
measurements should be performed. 
Therefore, the observation of inverse coherence peaks is the
evidence of the 3D odd-parity fully-gapped topological superconductivity. 

\begin{table}[htdp]
\caption{
\label{tab:classification}
Parity classification of time-reversal-invariant 3D
 superconductors. }
\begin{ruledtabular}
\begin{tabular}{ccccc}
Case &  Orbital & Spatial inversion & Fermi-surface topology & 1D $Z_{2}$ \\
I & +1 &+1 &+1 & -\\
I\!I & +1 & +1 &-1 & -\\
I\!I\!I & +1 & -1 & +1 & even \\
I\!V & +1 & -1 & -1 & odd \\
V (forbidden)& -1 &+1 & +1 & -\\
V\!I (forbidden)& -1 &+1 & -1 & -\\
V\!I\!I & -1 &-1 & +1 & even\\
V\!I\!I\!I & -1 &-1 & -1 & odd
\end{tabular}
\end{ruledtabular} 
\end{table}%


\begin{thebibliography}{99}
\bibitem{note}
Our choice is related to the Dirac representation $\gamma^{i}_{\rm D}$ by 
$\gamma^{i} = \hat{P} \gamma^{i}_{\rm D} \hat{P}^{\dagger}$ with $\hat{P}_{11} = s_{0}$, $\hat{P}_{12} = -i s_{z}$,$\hat{P}_{21} = s_{0}$, and $\hat{P}_{22} = i s_{z}$. $s_{0}$ is a unit matrix in spin space.
\end{thebibliography}
\begin{thebibliography}{99}
\bibitem{note:orbital_parity}
Using the linear map of partial transpose with respect to orbital degrees, 
we find for each 
\(
\hat{\Delta}^{A} \,(\propto \Gamma^{A}\gamma^{2}\gamma^{5})
\) that 
\(
2p_{\rm o} = 1 + p - m_{y} + m_{z}
\), with $m_{y,z} = \pm 1$. 
This identity indicates that odd orbital parity 
is not compatible with even parity.
\end{thebibliography}
\end{document}